\documentclass[seceq,preprint]{ptptex}

\usepackage{graphicx}
\usepackage{wrapft}
\usepackage{amsmath}
\notypesetlogo

\title{%        %You can use \\ for explicit line-break.
Numerical Calculation of Schwinger-Dyson Equation with Momentum-Dependent Gauge Parameter at Finite Temperature%
}

\author{%       %Use \scshape for the family name.
Shuji \textsc{Sasagawa} and Hidekazu \textsc{Tanaka}%
}

\inst{%     %Affiliation, neglected when [addenda] or [errata].
Department of Physics, Rikkyo University, Tokyo 171-8501, Japan
}

\recdate{October 20, 2009; Revised February 9, 2010}%            %Editorial Office will fill in this.

\abst{%         %This abstract is neglected when [addenda] or [errata].
Chiral symmetry at finite temperature is studied using the Schwinger-Dyson equation.
 We calculate numerically the critical temperature using the Schwinger-Dyson equation with the gauge parameter that depends on an external momentum.
 The critical temperature obtained by this method is similar to that with the Landau gauge and wave function renormalization constant 1.
 Moreover, the gauge invariance in the ladder approximation is examined using our method.%
}

\begin{document}

\maketitle

\section{Introduction}
In quantum chromodynamics (QCD), chiral symmetry is approximately realized at the lagrangian level, and it is broken by the strong interaction at zero temperature. On the other hand, at finite temperature and density, the broken symmetries are restored. Therefore, one expects that finite temperature and density QCD have various phases. For example, the quark-gluon plasma (QGP), where quarks and gluons are deconfined has been observed at the Relativistic Heavy Ion Collider (RHIC).\cite{rf:1}\ \ The QGP is examined at the Large Hadron Collider (LHC)\cite{rf:2}.\par
To study the chiral symmetry breaking, we need a nonperturbative treatment due to the strong interaction. As a theoretical approach to understand the phase structure of QCD, the lattice QCD simulation is a powerful method.\cite{rf:3}\ \ The QCD phase structure at finite temperature is extensively studied by this method. The lattice QCD simulation derives the results that are consistent with the experiment at RHIC.\cite{rf:4}\par
Another theoretical approach is the Schwinger-Dyson equation (SDE) method.\cite{rf:5}\ \ The SDE is a valid method for both finite temperature and density. The lattice QCD for large chemical potentials is still inadequate. Moreover, the numerical analysis of the SDE does not require a large-scale computer like in the case of lattice simulation. When we solve the SDE, we need appropriate approximations, because the SDE is a group of infinitely coupled equations.\par
Although the ladder approximation is usually used to solve the SDE, the SDE with this approximation depends on a gauge parameter. Owing to the gauge parameter dependence of the SDE, observable quantities, which should essentially be gauge-parameter-independent, depend on a gauge parameter. However, at zero temperature and density, the ladder approximation Ward-Takahashi identity (WTI) is guaranteed by choosing the Landau gauge. Thus, the Landau gauge is used at zero temperature and density.\cite{rf:6}\par
By contrast, at finite temperature and/or density, the ladder approximation WTI is not guaranteed trivially.\cite{rf:7}\ \ For this reason, there are no gauge parameters that have a clear advantage at finite temperature and/or density. The Landau gauge is adopted from the analogy of zero temperature\cite{rf:8,rf:9}\ . On the other hand, the Feynman gauge is adopted for convenience.\cite{rf:7}\par
The purpose of this paper is to perform the numerical calculation of the ladder approximation SDE to satisfy the WTI at finite temperature. For this purpose, we employ the method with a gauge parameter that depends on an external momentum. In this method, the WTI is satisfied using this functional gauge parameter. Such an idea, in which a gauge parameter is treated as a function, was studied at zero temperature\cite{rf:10} and was used in QED with real time formalism at finite temperature.\cite{rf:11}\ \ (The formulation at finite temperature is shown, e.g., in Ref.\ 12).) However, the numerical calculation of the SDE in real time formalism uses further approximation (IE approximation\cite{rf:13}) in addition to the ladder approximation. Moreover, the results obtained with real time formalism using IE approximation at the zero-temperature limit do not correspond to zero temperature.\cite{rf:14}$\ \ $Thus, the present SDE in real time formalism is insufficient. Hence, in this paper, we use the method with the functional gauge parameter in QCD with imaginary time formalism. Since the numerical calculation of the SDE in imaginary time formalism need not use other approximations, this formalism is more reliable.\par
The paper is organized as follows. In $\S$\ref{sec:sde}, we review the zero-temperature and finite-temperature SDE. In $\S$\ref{sec:cal}, we provide numerical results with the functional gauge parameter. The numerical method for solving the SDE is iteration. Then, we calculate the critical temperature. A summary and discussion are found in $\S$4.

\section{Schwinger-Dyson equation}\label{sec:sde}
\subsection{Zero-temperature SDE}\label{sec:1} 
The SDE is derived using the CJT effective potential.\cite{rf:15}\ \ The SDE for quark is given by\\
\begin{equation}
G^{-1}(p)=S^{-1}(p)-ig^{2}C_{2}\displaystyle \int\frac{d^{4}q}{(2\pi)^{4}}\gamma_{\mu}D^{\mu\nu}(p-q)G(q)\Gamma_{\nu}(p,q),\\[0.28cm]
\end{equation}
where $C_{2}$ is the Casimir operator, $S(p)$ is the free quark propagator, $D^{\mu\nu}$ is the exact gluon propagator, $\Gamma_{\nu}$ is the exact quark-gluon-quark vertex, and $G(p)$ is the exact quark propagator,\\
\begin{equation}
G(p)=\displaystyle \frac{1}{A(p)\gamma_{\mu}p^{\mu}-B(p)}.\label{eq:G1}\\[0.28cm]
\end{equation}
We use the chiral limit for the free quark propagator. In the ladder approximation\cite{rf:5}, the SDE is written as\\
\begin{equation}
G^{-1}(p)=S^{-1}(p)-ig^{2}C_{2}\displaystyle \int\frac{d^{4}q}{(2\pi)^{4}}\gamma_{\mu}D_{0}^{\mu\nu}(p-q)G(q)\gamma_{\nu},\label{eq:sde}\\[0.28cm]
\end{equation}
where $D_{0}^{\mu\nu}$ is the free gluon propagator,

\[
D_{0}^{\mu\nu}(k)=\frac{-g^{\mu\nu}+k^{\mu}k^{\nu}/k^{2}}{k^{2}}-\xi\frac{k^{\mu}k^{\nu}}{k^{4}}.\\[0.28cm]
\]
Here, $\xi$ is a gauge parameter. From this equation, the SDE is divided into coupled equations for scalar functions $A(p)$ and $B(p)$.\par
In QCD, we use the improved ladder approximation\cite{rf:16} in which a coupling constant is replaced by the running coupling constant $g(p^{2},q^{2})$. The running coupling is included in a momentum integral.\par
After performing angular integral,\cite{rf:16,rf:17}$\\$
\begin{subequations}\begin{equation}
A(l)=1+\displaystyle \frac{g^{2}(l)C_{2}\xi}{16\pi^{2}l^{2}}\int_{0}^{l}ds\frac{s^{2}A(s)}{A^{2}(s)s+B^{2}(s)}+\frac{C_{2}\xi}{16\pi^{2}}\int_{l}^{\infty}ds\frac{g^{2}(s)A(s)}{A^{2}(s)s+B^{2}(s)},\\[0.28cm]
\end{equation}
\begin{equation}
B(l)=\displaystyle \frac{g^{2}(l)C_{2}(3+\xi)}{16\pi^{2}l}\int_{0}^{l}ds\frac{sB(s)}{A^{2}(s)s+B^{2}(s)}+\frac{C_{2}(3+\xi)}{16\pi^{2}}\int_{l}^{\infty}ds\frac{g^{2}(s)B(s)}{A^{2}(s)s+B^{2}(s)}.\\[0.28cm]
\end{equation}
\end{subequations}
where $l=p_{E}^{2}$ and $s=q_{E}^{2}$, the index $E$ shows four vectors in Euclidean space. In the ladder approximation, the wave function renormalization constant must be unity to satisfy the WTI, that is, $A(l)=1$. This equation for $A(l)$ shows that $A(l)=1$ if the Landau gauge $\xi=0$ is adopted. Thus, the WTI is satisfied using the Landau gauge at zero temperature.\par
Although the WTI is extended to the Slavnov-Taylor identities in QCD, the Slavnov-Taylor identities become the WTI type to omit ghost-quark scattering kernel.\cite{rf:18}\ \ Since the ladder approximation corresponds to this situation, we use QCD with the WTI type.\par

\subsection{Finite-temperature SDE}
The Feynman rules in imaginary time formalism are summarized as follows:

\[
\mathrm{free}\ \mathrm{q}\mathrm{u}\mathrm{a}\mathrm{r}\mathrm{k}\ \mathrm{p}\mathrm{r}\mathrm{o}\mathrm{p}\mathrm{a}\mathrm{g}\mathrm{a}\mathrm{t}\mathrm{o}\mathrm{r}:\ S(p)=\frac{-1}{\gamma_{\mu}p^{\mu}-m}\ \ \Big(p_{0}=i\omega_{n}=2\pi iT\Big(n+\frac{1}{2}\Big)\Big)\\
\]

\[
\mathrm{free}\ \mathrm{g}\mathrm{l}\mathrm{u}\mathrm{o}\mathrm{n}\ \mathrm{p}\mathrm{r}\mathrm{o}\mathrm{p}\mathrm{a}\mathrm{g}\mathrm{a}\mathrm{t}\mathrm{o}\mathrm{r}:\ D_{0}^{\mu\nu}(k)=\frac{g^{\mu\nu}-k^{\mu}k^{\nu}/k^{2}}{k^{2}}+\xi\frac{k^{\mu}k^{\nu}}{k^{4}}\ \ (k_{0}=i\omega_{l}=2\pi iTl)\\
\]

\[
(2\pi)^{4}\delta^{4}(p+\cdots)\ \Rightarrow\ -\frac{i}{T}(2\pi)^{3}\delta_{n,\cdots}\delta^{3}(\boldsymbol{p}+\cdots),\\
\]

\[
\int\frac{d^{4}p}{(2\pi)^{4}}\ \Rightarrow\ iT\sum_{n}\int\frac{d^{3}p}{(2\pi)^{3}}.\\
\]
Here, $\omega_{n}$ is the Matsubara frequency. By this replacement, the improved ladder approximation SDE for quark in imaginary time formalism is given by$\\$
\begin{equation}
G^{-1}(p)=S^{-1}(p)-C_{2}T\displaystyle \sum_{m}\int\frac{d^{3}q}{(2\pi)^{3}}g^{2}(-p^{2},-q^{2})\gamma_{\mu}D_{0}^{\mu\nu}(p-q)G(q)\gamma_{\nu},\\
\end{equation}
where $p_{\mu}=(2i\pi T(n+1/2),p_{i})$ and $q_{\mu}=(2i\pi T(m+1/2),q_{i})$. If we use the exact quark propagator form, Eq.\ (\ref{eq:G1}), the coupled equations for $A_{n}^{\prime}(x)$ and $B_{n}(x)$ are written as\footnote{Since a propagator at finite temperature is Euclidean, Eq.(\ref{eq:G1}) is replaced by $G_{n}(x)=-1/(A_{n}^{\prime}(x)p_{\mu}\gamma^{\mu}-B_{n}(x))$.}\par
\begin{subequations}\begin{equation}
A_{n}^{\prime}(x)=1-\displaystyle \frac{C_{2}T}{p^{2}}\frac{1}{8\pi^{2}x}\sum_{m}\int_{0}^{\infty}dyg^{2}(-p^{2},-q^{2})\frac{yA_{m}^{\prime}(y)[L_{1}+L_{2}+\xi(L_{1}-L_{2})]}{A_{m}^{\prime 2}(y)q^{2}-B_{m}^{2}(y)},\label{eq:A1}\\[0.28cm]
\end{equation}
\begin{equation}
B_{n}(x)=-C_{2}T\displaystyle \frac{(3+\xi)}{8\pi^{2}x}\sum_{m}\int_{0}^{\Lambda}dy\frac{g^{2}(-p^{2},-q^{2})yB_{m}(y)}{A_{m}^{\prime 2}(y)q^{2}-B_{m}^{2}(y)}\log\frac{(p_{0}-q_{0})^{2}-(x+y)^{2}}{(p_{0}-q_{0})^{2}-(x-y)^{2}},\label{eq:A2}
\end{equation}
\end{subequations}
where, $x=|\boldsymbol{p}|$ and $y=|\boldsymbol{q}|$. Alternatively, if we use the general form of the quark propagator at finite temperature,$\\$
\begin{equation}
G_{n}(x)=\displaystyle \frac{-1}{C_{n}(x)\gamma_{0}p_{0}+A_{n}(x)\gamma_{i}p^{i}-B_{n}(x)},\\[0.28cm]
\end{equation}
the coupled equations for $A_{n}(x),\ B_{n}(x)$, and $C_{n}(x)$ are$\\$
\begin{subequations}\begin{equation}\\[0.35cm]
\displaystyle \begin{split}C_{n}(x)=1&+\displaystyle \frac{C_{2}T}{8\pi^{2}p_{0}x}\sum_{m}\int_{0}^{\infty}dyg^{2}(-p^{2},-q^{2})y\\&\displaystyle \times\frac{-C_{m}(y)(I_{1}+I_{2})-A_{m}(y)I_{3}+\xi(C_{m}(y)I_{2}+A_{m}(y)I_{3})}{C_{m}^{2}(y)q_{0}^{2}-A_{m}^{2}(y)y^{2}-B_{m}^{2}(y)},\end{split}\label{eq:gene1}
\end{equation}
\begin{equation}\\[0.35cm]
\displaystyle \begin{split}A_{n}(x)=1&-\displaystyle \frac{C_{2}T}{8\pi^{2}x^{3}}\sum_{m}\int_{0}^{\infty}dyg^{2}(-p^{2},-q^{2})y\\\times&\displaystyle \frac{-C_{m}(y)H_{1}+A_{m}(y)(H_{2}-H_{3})+\xi(C_{m}(y)H_{1}+A_{m}(y)H_{3})}{C_{m}^{2}(y)q_{0}^{2}-A_{m}^{2}(y)y^{2}-B_{m}^{2}(y)},\end{split}\label{eq:gene2}
\end{equation}
\begin{equation}
\displaystyle \begin{split}B_{n}(x)=-C_{2}T\frac{3+\xi}{8\pi^{2}x}\sum_{m}\int_{0}^{\infty}dy&\displaystyle \frac{g^{2}(-p^{2},-q^{2})yB_{m}(y)}{C_{m}^{2}(y)q_{0}^{2}-A_{m}^{2}(y)y^{2}-B_{m}^{2}(y)}\\&\hspace{4em}\displaystyle \times\log\frac{(p_{0}-q_{0})^{2}-(x+y)^{2}}{(p_{0}-q_{0})^{2}-(x-y)^{2}}.\end{split}\label{eq:gene3}
\end{equation}\end{subequations}
The explicit expressions of $L,\ I$, and $H$ are given in Appendix A. Equations (\ref{eq:gene1})--(\ref{eq:gene3}) are the general coupled equations derived from the ladder approximation SDE at finite temperature. Those four scalar functions have the relation for $n$, e.g., $B_{n}(x)=B_{-n-1}(x).$\par
In this paper, we use the following form for the running coupling:\cite{rf:19}$\\$
\begin{equation}
g^{2}(-p^{2},-q^{2})=\displaystyle \frac{48\pi^{2}}{11N_{c}-2N_{f}}\times\left\{\begin{array}{l}
\frac{1}{t}\ \ \hspace{8.8em},\ \ t_{F}<t,\\
\frac{1}{t_{F}}+\frac{(t_{F}-t_{C})^{2}-(t-t_{C})^{2}}{2t_{F}^{2}(t_{F}-t_{C})}\ \ \ ,\ \ t_{C}<t<t_{F},\\
\frac{1}{t_{F}}+\frac{t_{F}-t_{C}}{2t_{F}^{2}}\ \ \hspace{4.6em},\ \ t<t_{C},
\end{array}\right.\label{eq:couple}\\
\end{equation}
where $t=\log[(-p^{2}-q^{2})/\Lambda_{qcd}^{2}],\ t_{C}=-2,\ t_{F}=0.5,\ \Lambda_{qcd}=592(\mathrm{MeV})$, and $N_{c}$ and $N_{f}$ are the numbers of colors and flavors respectively. Here, we use $N_{c}=3$ and $N_{f}=2.\ B_{n}(x)$ depends on the regularization parameter $t_{F}$. For parameters in running coupling, we use the parameters in Ref. 8) (see $\S$3.2).

\section{Results of numerical calculation}\label{sec:cal}
\subsection{Functional gauge parameter}
The WTI is satisfied by taking the Landau gauge at zero temperature, as shown in $\S$\ref{sec:1}. By contrast, the WTI is not satisfied using a constant gauge parameter at finite temperature. In fact, $A_{n}^{\prime}(x),\ C_{n}(x)$, and $A_{n}(x)$ are not unity at finite temperature even if one takes the Landau gauge (see Figs.\ \ref{fig:CA3} and \ref{fig:CA2}). Hence, we assume that the gauge parameter is a function depending on the external momentum.\par
We divide Eqs.\ (\ref{eq:A1}), (\ref{eq:gene1}), and (\ref{eq:gene2}) into the gauge parameter term and no gauge parameter term, e.g.,$\\$
\begin{equation}
C_{n}(x)=1+\xi X_{n}(x)+Y_{n}(x).\label{eq:base}\\
\end{equation}
In this equation, we treat $\xi$ as a function dependent on the external momentum $\omega_{n},\ x$. Since a gauge parameter exists in the gluon propagator, it should fundamentally depend on a momentum of gluon (see Eq.\ (\ref{eq:sde})). However, for simplicity, we use the functional gauge parameter that depends on only the external momentum. (Although we do not write in terms, $\xi_{n}(x)$ clearly depends on temperature.) From this assumption, we configured that the gauge parameter must satisfy $\xi_{n}(x)X_{n}(x)+Y_{n}(x)=0$. By this method, we can perform numerical calculation satisfying the WTI.\par
Note that there is a problem for Eqs.\ (\ref{eq:gene1}) and (\ref{eq:gene2}). It is the fact that the temperature dependences of $C_{n}(x)$ and $A_{n}(x)$ are different. Owing to this property, it is difficult to make $C_{n}(x)=1$ and $A_{n}(x)=1$ at the same time. However, $C_{n}(x)$ and $A_{n}(x)$ have near values at around the critical temperature. Hence, in this important region, we can simultaneously make $C_{n}(x)\simeq 1$ and $A_{n}(x)\simeq 1$. We operate $C_{n}(x)$ and $A_{n}(x)$ so that those may approach $1$ as much as possible.

\subsection{Numerical calculation}
Although an effective potential is needed to study the phase transition, $B_{n}(x)=0$ directry denotes a critical point at finite temperature\cite{rf:9}. Thus, we focus our attention on the critical point at $B_{n}(x)=0$.\par
Note that our numerical calculation has an error due to the number of flavors in the running coupling. In Ref.\ 8), the running coupling with $N_{f}=3$ was used. However, we took $N_{f}=2$. This difference in $N_{f}$ in the running coupling with the same $\Lambda_{qcd}$ results in the difference of about $2$ (MeV) for the pion decay constant. However, in this paper, we are designed to search for the existence of a solution satisfying the WTI and the shift of the critical temperature. Thus, this difference in the pion decay constant is not important for that purpose of the study. (In addition, this difference might be within the range of numerical error.) The critical temperature has the error range of about $\pm 3$ (MeV) at least by ignoring this.\par
We calculated the following four cases:$\\$
\begin{description}
\item[(I):]\ coupled equations, $A_{n}^{\prime}(x)$ and $B_{n}(x)$ with the Landau gauge,
\item[(II):]\ coupled equations, $C_{n}(x),A_{n}(x)$ and $B_{n}(x)$ with the Landau gauge,
\item[(Ia):]\ coupled equations, $A_{n}^{\prime}(x)$ and $B_{n}(x)$ with the functional gauge parameter $\xi_{n}^{\prime}(x)$, and
\item[(IIa):]\ coupled equations, $C_{n}(x),A_{n}(x)$ and $B_{n}(x)$ with the functional gauge parameter $\xi_{n}(x)$.\\
\end{description}
To solve the SDE, we employed the iteration method. It starts as a constant or a trial function, and repeats until a value is converged. First, we tried the trial function like zero temperature (see, e.g., Refs.\ 17) and 19)). The trial function has the zero temperature form for $x$ at $n=0,\ -1$; others are small constant values. However, there is no difference between a constant and the trial function except the convergence. Thus, we used a constant.\par
The range of summation $m=-10\sim 9$ is sufficiently large as a truncation point for summation in the case of $A_{n}^{\prime}(x)=1$. In contrast, in (I) and (II), since the range of $-10\sim 9$ is insufficient (especially in the calculation of the critical temperature), the summation is necessary to take above $m=-40\sim 39$.\par
Since $C_{n}(x)$ and $A_{n}(x)$ simultaneously do not make $C_{n}(x),A_{n}(x)\simeq 1$ below the critical temperature, (IIa) is used only for determining the critical temperature. Then, we fix the lowest value of $C_{n}(x)$ as $0.94$ and the highest value as $1$, because it is possible to limit the difference from $1$ to about 0.05 at around the critical temperature.\par
If one fixes $C_{n}(x)=1$ exactly at all regions, $A_{n}(x)$ is further away from $1$, compared with the case of $C_{n}(x)=0.94\sim 1$. Thus, we do not fix $C_{n}(x)=1$ exactly even around the critical temperature. We fix $A_{n}^{\prime}(x)=1$ for (Ia) and limit $C_{n}(x)=0.94\sim 1$ for (IIa).\par
On the other hand, we assume that $C_{n}(x)=A_{n}(x)=1$ in (IIa) has a problem, because $C(p)\neq A(p)$ ($p_{0}$ is continuous) in real time generates plasminos for fermion.\cite{rf:20}\ \ From this viewpoint, $C_{n}(x)\simeq A_{n}(x)\simeq 1$ might be valid as an approximation (see also $\S$4).
\par
\begin{figure}[b]
\begin{tabular}{cc}
\begin{minipage}{0.5\hsize}
\begin{center}
\includegraphics[width=60mm]{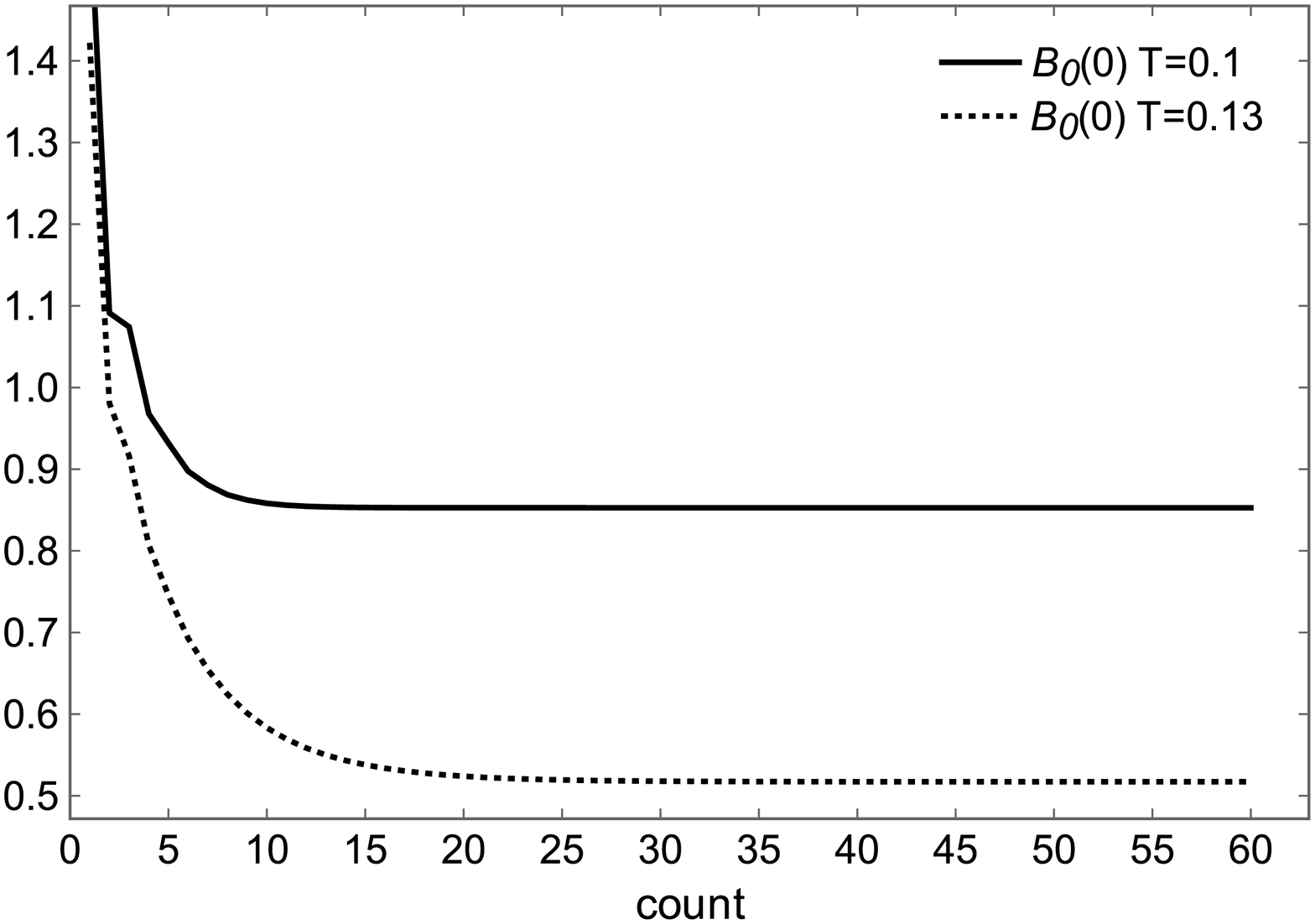}
\caption{Convergence of $B_{0}(0)$ at $T=0.1,0.13$ \newline(GeV).}
\label{fig:converB}
\end{center}
\end{minipage}
\begin{minipage}{0.5\hsize}
\begin{center}
\includegraphics[width=60mm]{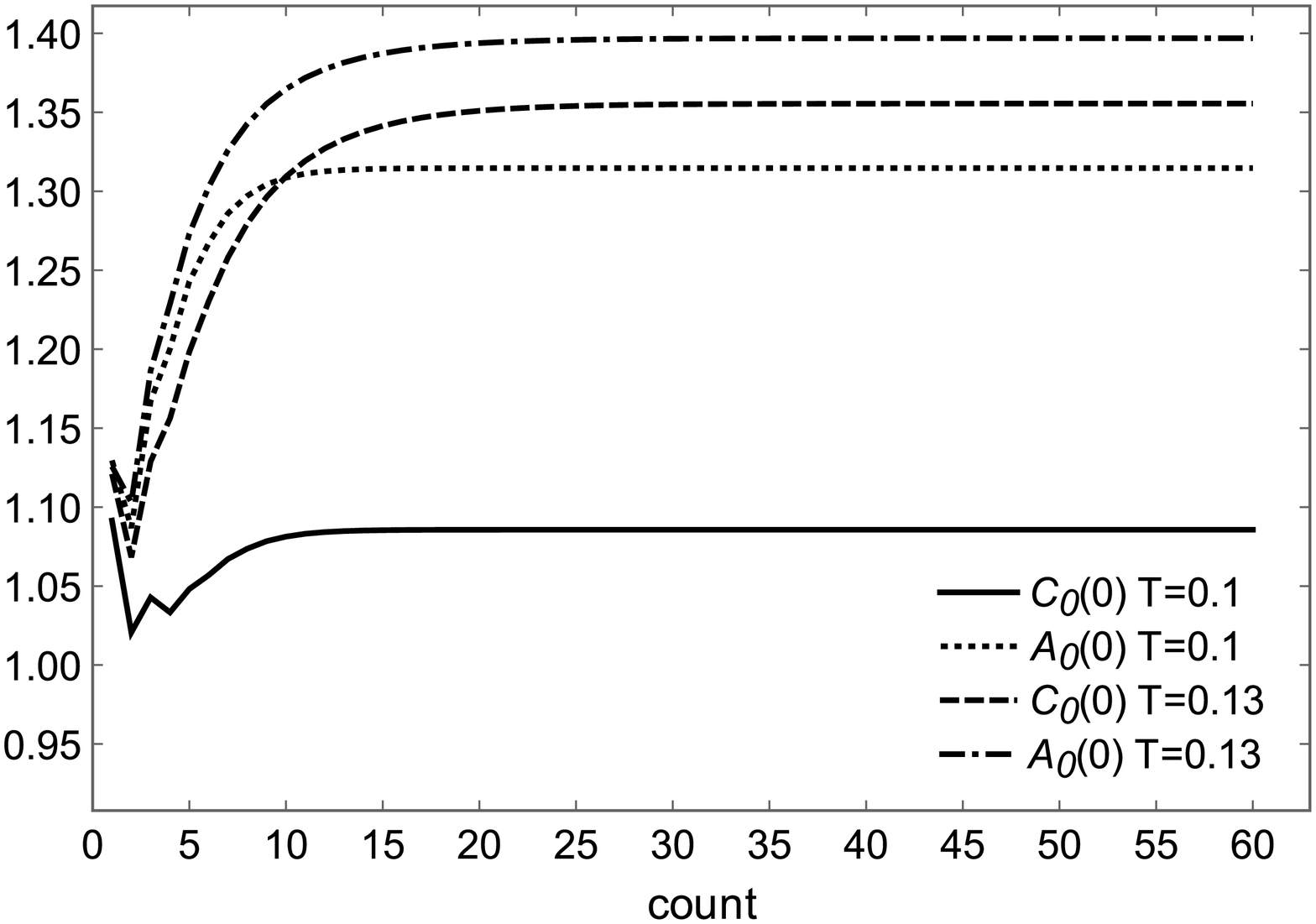}
\caption{Convergence of $C_{0}(0),A_{0}(0)$ at $T=0.1$,\newline$0.13$ (GeV).}
\label{fig:converA}
\end{center}
\end{minipage}
\end{tabular}
\end{figure}

The convergence for $B_{0}(0),\ C_{0}(0)$ and $A_{0}(0)$ in (II) is shown in Figs.\ \ref{fig:converB} and \ref{fig:converA}. The convergence is worse around the critical temperature. Other cases also have the same convergence.\par

\subsection{Results of (I) and (II)}
\begin{figure}[t]
\begin{tabular}{ll}
\begin{minipage}{0.5\hsize}
\begin{center}
\includegraphics[width=65mm]{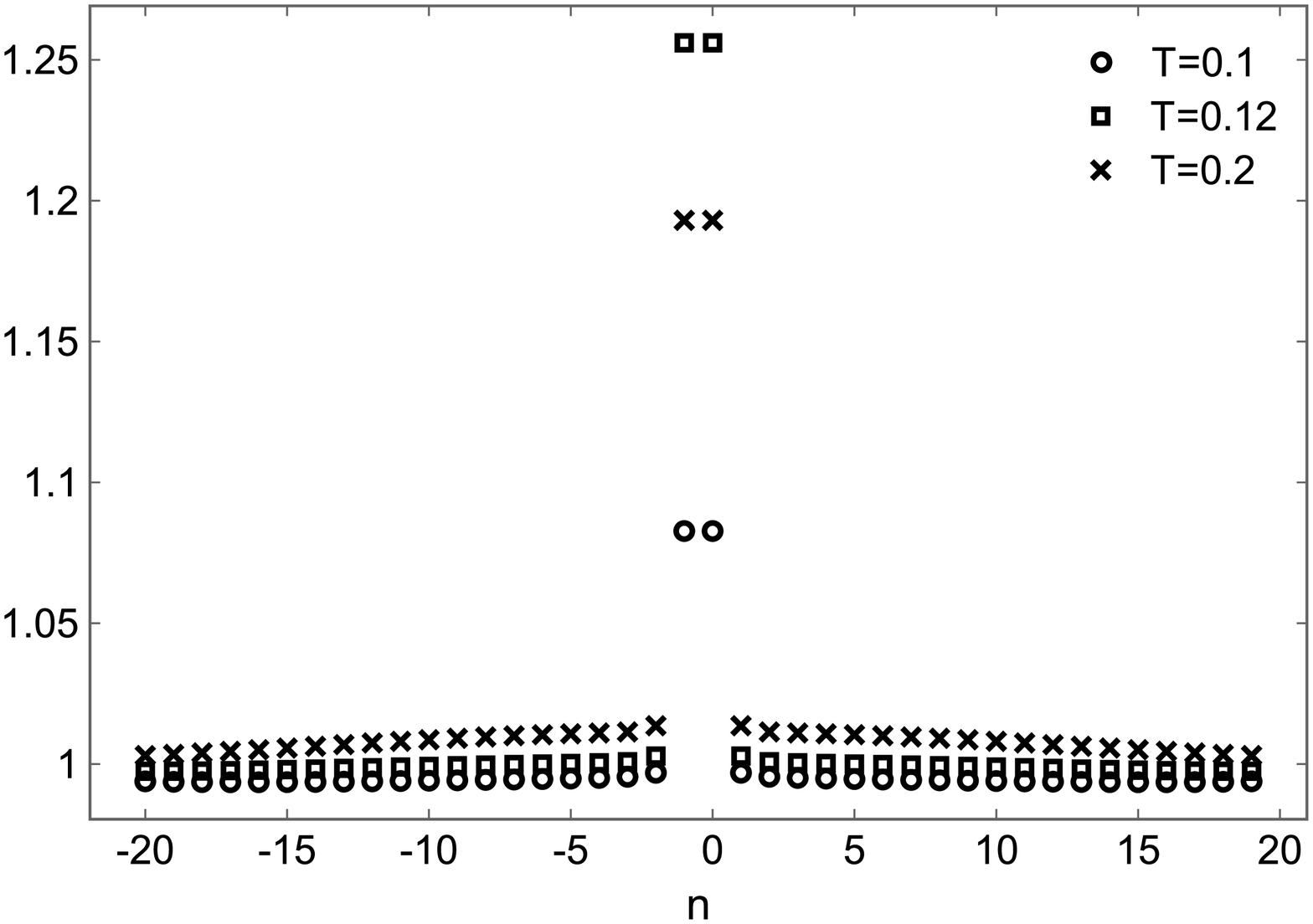}
\caption{Behavior of $A_{n}^{\prime}(0)$ for $n$ at $T=0.1$,\newline$0.12,0.2$ (GeV).}
\label{fig:CA3}
\end{center}
\end{minipage}
\begin{minipage}{0.5\hsize}
\begin{center}
\includegraphics[width=65mm]{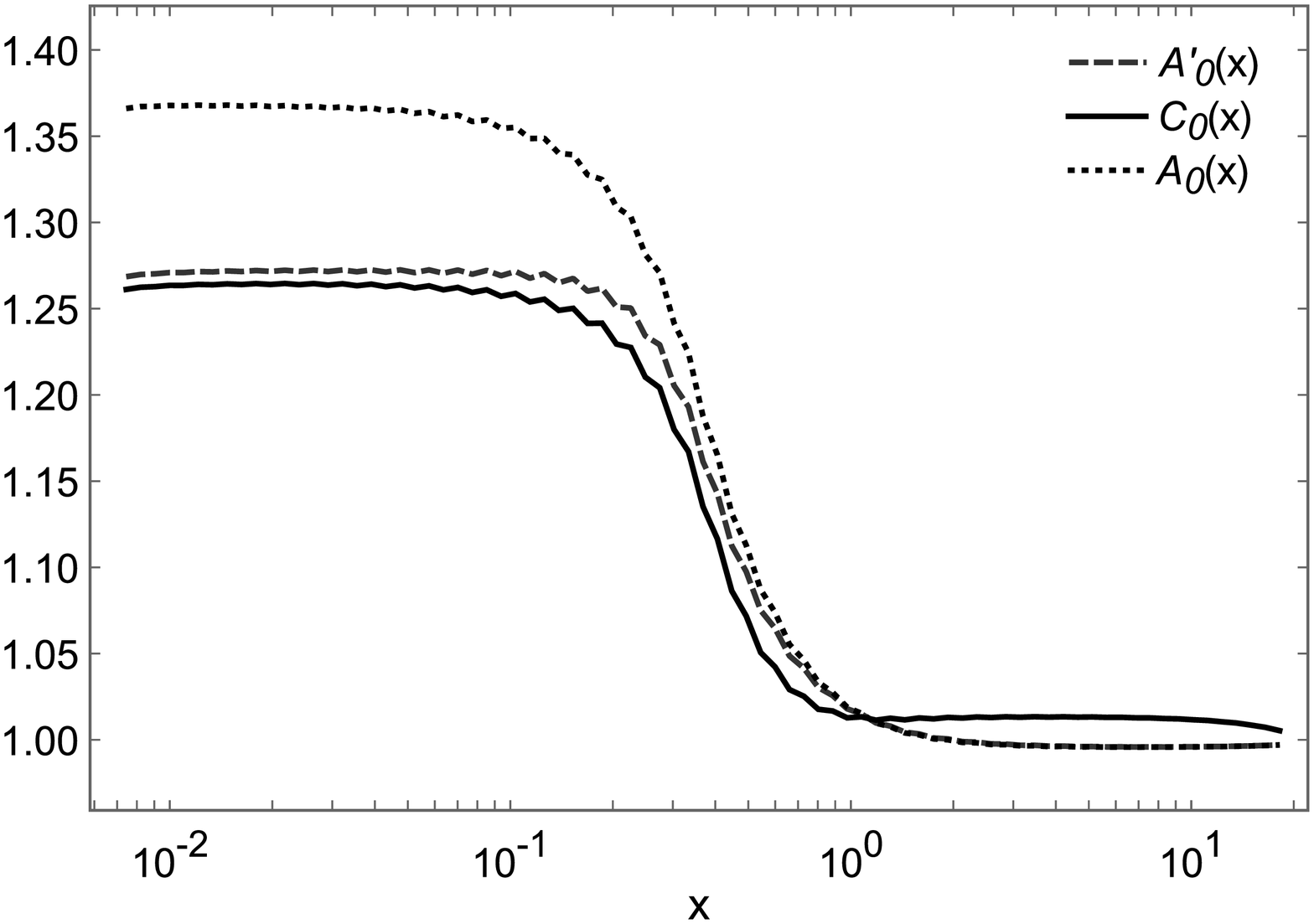}
\caption{$x$ dependences of $A_{0}^{\prime}(x)$ and $C_{0}(x),\ A_{0}(x)$ at $T=0.12$ (GeV).}
\label{fig:CA2}
\end{center}
\end{minipage}
\end{tabular}
\end{figure}

\begin{figure}[t]
\centerline{\includegraphics[width=9cm]{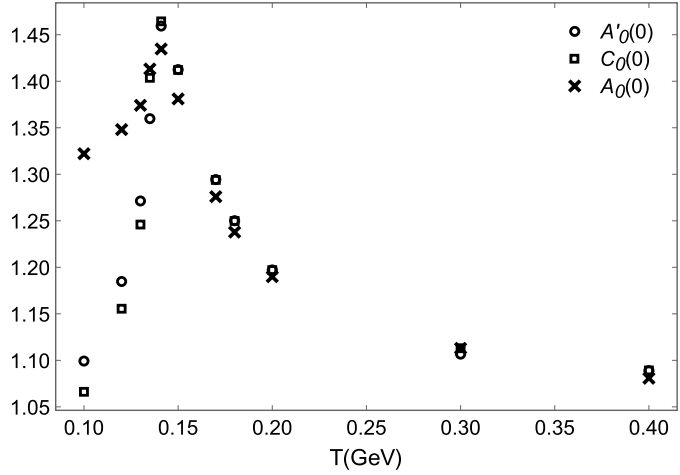}}
\caption{Temperature dependences of $A_{0}^{\prime}(0)$ and $C_{0}(0),\ A_{0}(0)$. Those have a peak around the critical temperature.}
\label{fig:CA1}
\end{figure}

The typical $x$ and $n$ dependences of $A_{n}^{\prime}(x)$ and $C_{n}(x),\ A_{n}(x)$ are shown in Figs.\ \ref{fig:CA3} and \ref{fig:CA2}. (Figure \ref{fig:CA3} shows only the case of $A_{n}^{\prime}(0)$. The behavior for $C_{n}(0),\ A_{n}(0)$ is much the same.) The temperature dependences of $A_{0}^{\prime}(0)$ and $C_{0}(0),\ A_{0}(0)$ are also shown in Fig.\ \ref{fig:CA1}. Those results actually show $A_{n}^{\prime}(x)\neq 1$ and $C_{n}(x)\neq 1,\ A_{n}(x)\neq 1$. In particular, $A_{0}^{\prime}(0)$ and $C_{0}(0),\ A_{0}(0)$ shift from $1$ mostly in the vicinity of the critical temperature. (The critical temperature is shown in Fig.\ \ref{fig:B}.) Thus, $A_{n}^{\prime}(x)$ and $C_{n}(x),\ A_{n}(x)$ strongly contribute to the critical temperature. The effect of $A_{n}^{\prime}(x)$ or $C_{n}(x),\ A_{n}(x)$ on the critical temperature with $A_{n}^{\prime}(x)=1$ is estimated at about $30$ (MeV).\par

\subsection{Results of (Ia) and (IIa)}
\begin{figure}[t]
\begin{tabular}{cc}
\begin{minipage}{0.5\hsize}
\begin{center}
\includegraphics[width=65mm]{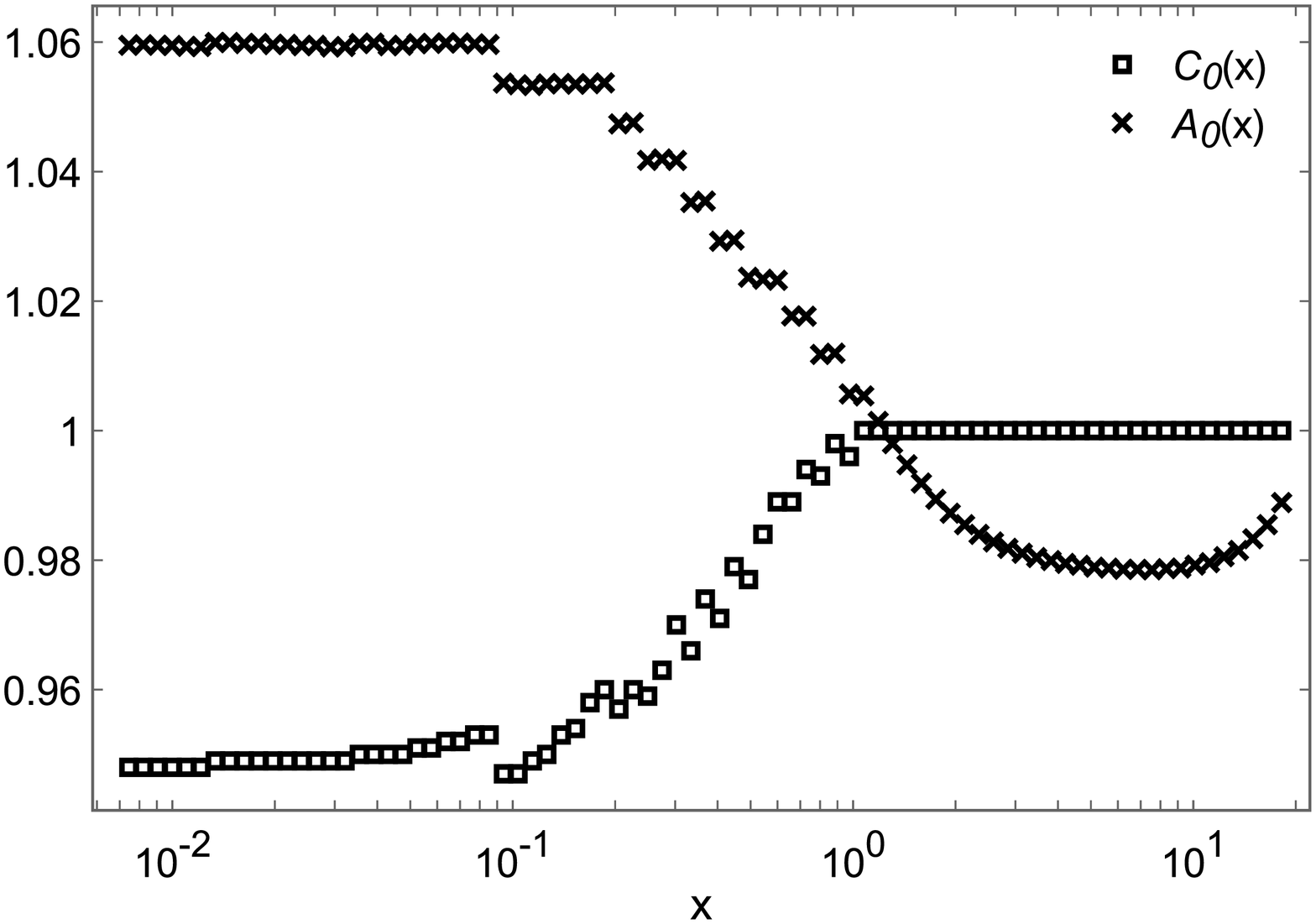}
\caption{$x$ dependence of $C_{0}(x),\ A_{0}(x)$ with \newline the functional gauge parameter at $T$\newline$=0.15$ (GeV).}
\label{fig:xiCA1}
\end{center}
\end{minipage}
\begin{minipage}{0.5\hsize}
\begin{center}
\includegraphics[width=65mm]{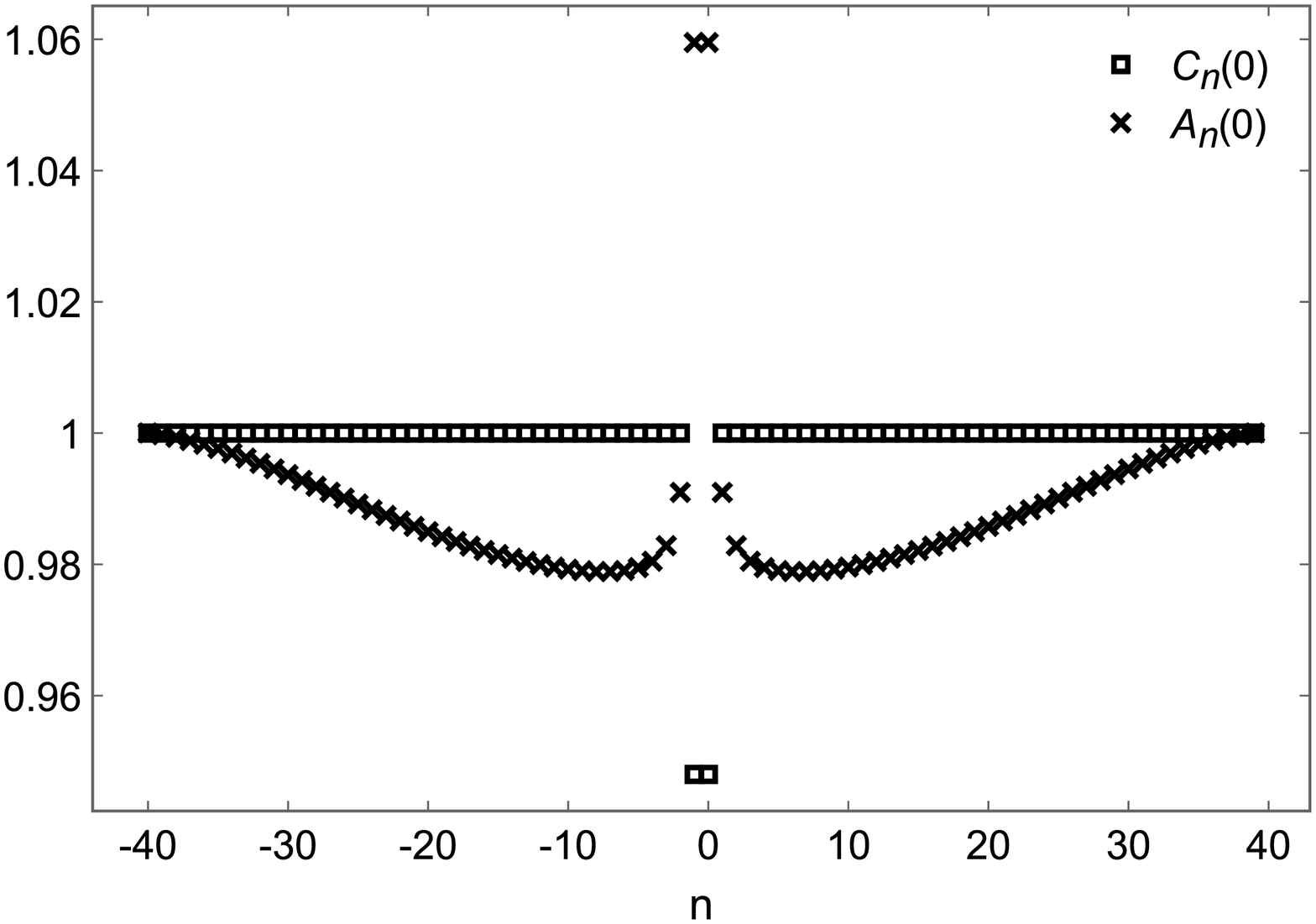}
\caption{$n$ dependence of $C_{n}(0),\ A_{n}(0)$ with the functional gauge parameter at $T=0.15$ (GeV).}
\label{fig:xiCA2}
\end{center}
\end{minipage}
\end{tabular}
\end{figure}

In (Ia), $A_{n}^{\prime}(x)$ is unity at all regions. In (IIa), the behaviors of $C_{n}(x)$ and $A_{n}(x)$ with the functional gauge parameter are shown in Figs.\ \ref{fig:xiCA1} and \ref{fig:xiCA2}. The dependence of $x$ for $n\neq 0$ is also a similar form. Owing to forcibly making the situation $C_{n}\simeq 1,\ n$ and $x$ dependences of $A_{n}(x)$ change from the case of (II). Nevertheless, since $n$ and $x$ dependences of $B_{n}(x)$ are unchanged in the result of numerical calculation ($B_{n}(x)$ is shown, e.g., Refs.$\ $8) and 9)), we expect that the properties of $B_{n}(x)$ are not lost.\par
\begin{figure}[t]
\centerline{\includegraphics[width=9cm]{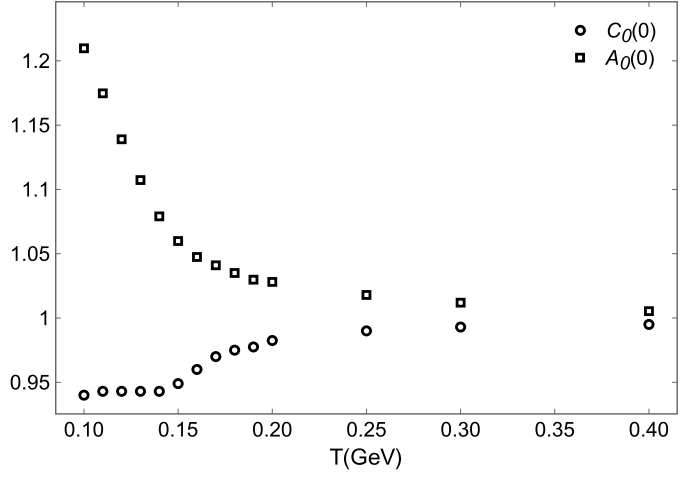}}
\caption{Temperature dependence of $C_{0}(0),\ A_{0}(0)$ with the functional gauge parameter.}
\label{fig:xiCA3}
\end{figure}

$C_{n}(x)$ and $A_{n}(x)$ have values that are closer to $1$ than the result with the Landau gauge, e.g., $C_{0}(0)=1.264$ and $A_{0}(0)=1.368$ in (II), $C_{0}(0)=0.943$ and $A_{0}(0)=1.139$ in (IIa) at $T=0.12$ (GeV). Moreover, $C_{n}(x)$ and $A_{n}(x)$ approach $1$ simultaneously above the critical temperature. ($C_{0}(0)$ and $A_{0}(0)$ are shown Fig.\ \ref{fig:xiCA3}. Other cases also have this feature). However, as mentioned in $\S$3.1, this method cannot achieve $C_{n}\simeq 1$ and $A_{n}\simeq 1$ simultaneously at low temperature.\par

\begin{figure}[t]
\begin{tabular}{cc}
\begin{minipage}{0.5\hsize}
\begin{center}
\includegraphics[width=65mm]{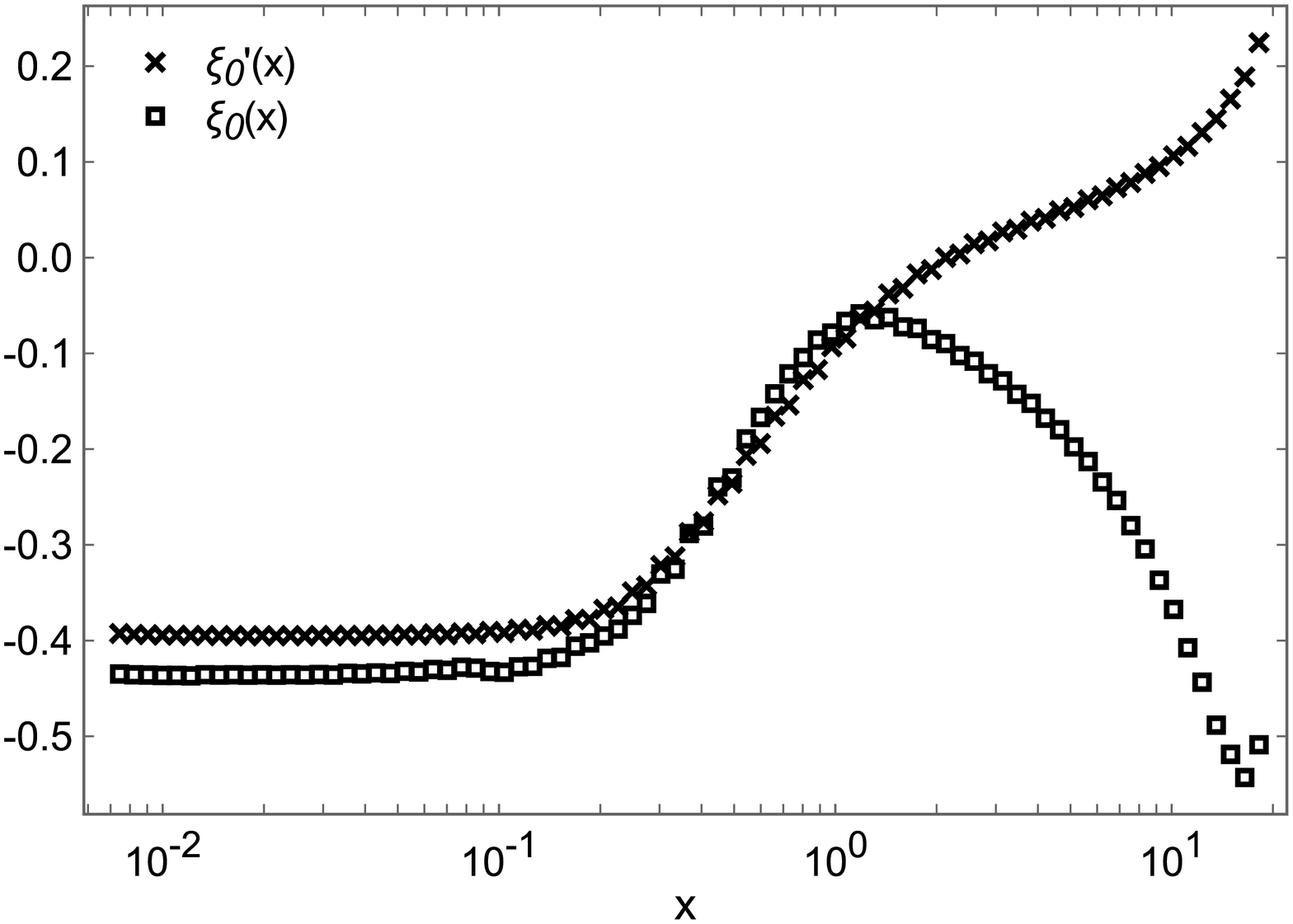}
\caption{$x$ dependence of $\xi_{0}(x)$ at $T=0.15$ (\newline GeV)}
\label{fig:xi1}
\end{center}
\end{minipage}
\begin{minipage}{0.5\hsize}
\begin{center}
\includegraphics[width=65mm]{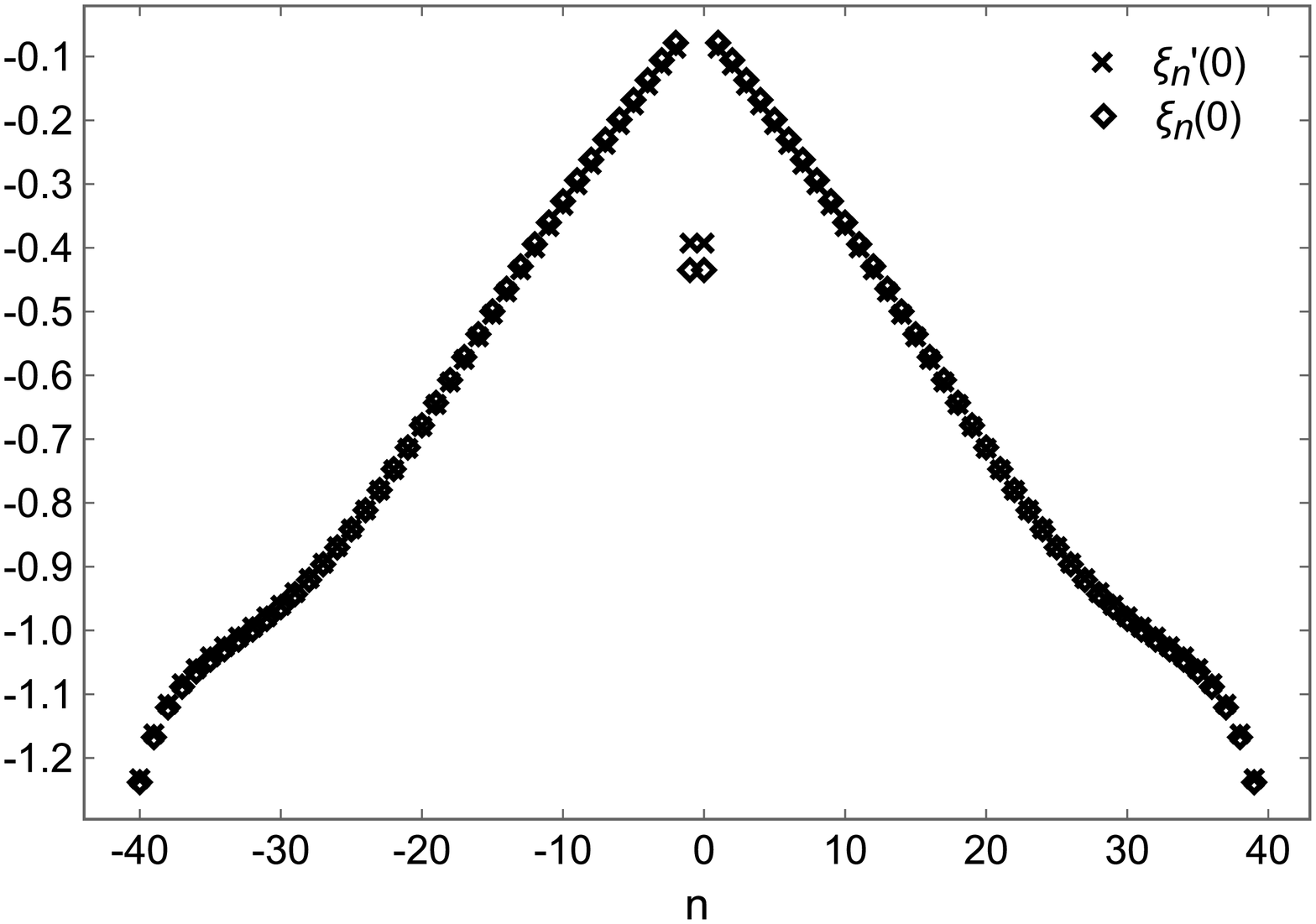}
\caption{$n$ dependence of $\xi_{n}(0)$ at $T=0.15$ (GeV)}
\label{fig:xi2}
\end{center}
\end{minipage}
\end{tabular}
\end{figure}

The functional gauge parameters $\xi_{n}^{\prime}(x)$ and $\xi_{n}(x)$ are shown in Figs.\ \ref{fig:xi1} and \ref{fig:xi2}. $\xi_{0,-1}^{\prime}(x)$ ($\xi_{0,-1}(x)$) has the most different value from the Landau gauge $\xi=0$.\par
$\xi_{n}^{\prime}(x)$ and $\xi_{n}(x)$ have similar values at around the critical temperature, because $A_{n}^{\prime}(x)$ resembles $C_{n}(x)$ to some degree (see Figs.\ \ref{fig:CA2} and \ref{fig:CA1}). However, there is a different behavior above $x=1$ (GeV) (Fig.\ \ref{fig:xi1}). This difference is understood from Eq.\ (\ref{eq:base}) and Fig.\ \ref{fig:CA2}. In Eq.\ (\ref{eq:base}), if $C_{n}(x)$ with the Landau gauge is below $1$ ($Y_{n}(x)$ is negative) and $X_{n}(x)$ is positive, $\xi_{n}(x)$ is positive. Similarly, if $C_{n}(x)$ with the Landau gauge is above $1$ and $X_{n}(x)$ is positive, $\xi_{n}(x)$ is negative etc. Therefore, the difference above $x=1$ (GeV) in Fig.\ \ref{fig:xi1} results from a sign of $X_{n}(x),\ C_{n}(x)-1$ and $A_{n}^{\prime}(x)-1$.\par
For large $|n|,\ \xi_{n}^{\prime}(x)$ and $\xi_{n}(x)$ have large values. For small $|n|$, they have large values at large $x$. Since $A_{n}^{\prime}(x)$ and $C_{n}(x),\ A_{n}(x)$ with the Landau gauge are about unity at that region, a gauge parameter does not contribute to $A_{n}^{\prime}(x)$ and $C_{n}(x),\ A_{n}(x)$ if a gauge parameter is not large. Hence, the large values of $\xi_{n}^{\prime}(x)$ and $\xi_{n}(x)$ at that region are not meaningful. Moreover, since $A_{n}^{\prime}(x)$ and $C_{n}(x),\ A_{n}(x)$ at that region are about unity, there is no problem in the use of the Landau gauge at that region. For this reason, it is possible to choose a gauge parameter as a step function depending on the external momentum. For example:$\\$
\[
\xi_{n}(x)=\alpha(T)\delta_{n,\lambda}\times\left\{\begin{array}{l}
\ 1\ \ \ x<\Lambda_{qcd}\\
\ 0\ \ \ x>\Lambda_{qcd}
\end{array}\right.\ ,\ \lambda=-2,-1,0,1,\\
\]
where $\alpha(T)$ is a function that has appropriate values for each temperature.

\begin{figure}[t]
\centerline{\includegraphics[width=9cm]{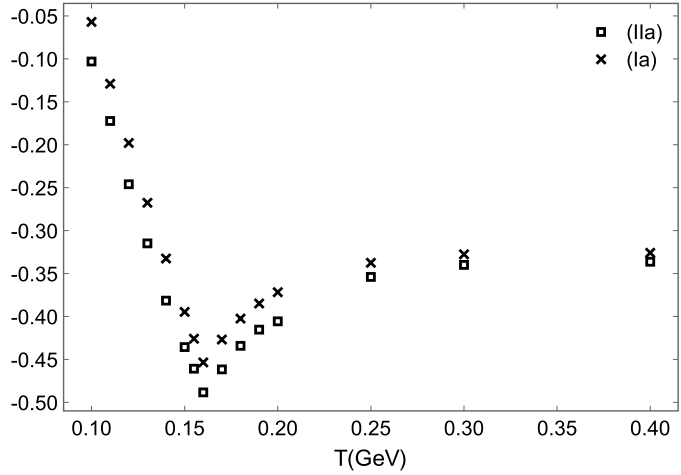}}
\caption{Temperature dependence of $\xi_{0}(0),\ \xi_{0}^{\prime}(0).\ \xi_{0}(0)$ and $\xi_{0}^{\prime}(0)$ have the downward peak corresponding to the largest values of $C_{0}(0)$ and $A_{0}^{\prime}(0)$ (see Fig.\ \ref{fig:CA1}).}
\label{fig:xi3}
\end{figure}

\subsection{Critical temperature}
\begin{figure}[t]
\centerline{\includegraphics[width=9.9cm]{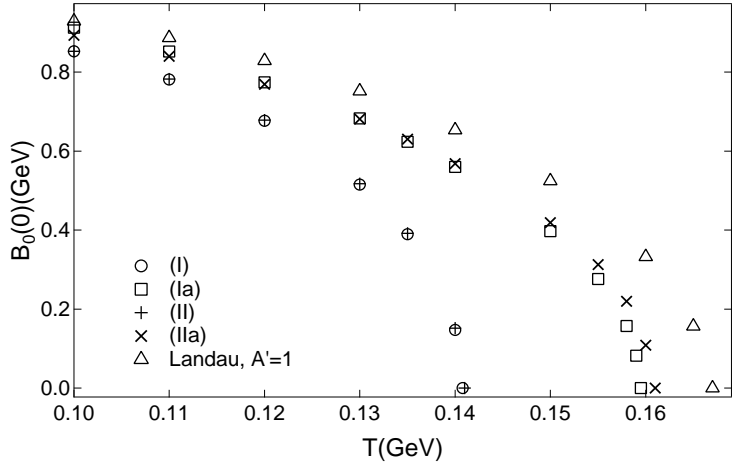}}
\caption{Temperature dependence of $B_{0}(0)$ in the cases of (I), (II), (Ia), (IIa), and the Landau gauge with $A_{n}^{\prime}(x)=1$.}
\label{fig:B}
\end{figure}

The temperature dependence of $B_{0}(0)$ is shown in Fig.\ \ref{fig:B}. (I) and (II) have similar temperature dependence and critical temperature. Their critical temperature is 141 (MeV). As the result, one finds that the difference in (I) and (II) hardly affects $B_{n}(x)$.\par
In Ref.\ 21), the case of $A_{n}^{\prime}(x)=1$ with Landau gauge and (II) was calculated with a different running coupling constant (Higashijima-Miransky type). This paper also showed that (II) has a lower critical temperature than in the case of $A_{n}^{\prime}(x)=1$ with Landau gauge. The difference in the running coupling constant is the method of infrared cutoff and, perhaps, the maximum value. These differences do not strongly affect the behavior of $B_{n}(x)$. Thus, the temperature dependence of $B_{n}(x)$ for these two cases is similar to this paper.\par
The critical temperature of (Ia) is 160 (MeV), (IIa) is 161 (MeV). Those also have almost the same value. The critical values of (Ia) and (IIa) are about 10 (MeV) lower than the result of the Landau gauge with $A_{n}^{\prime}(x)=1$.\par
On the other hand, since the choice of $C_{n}(x)$ contributes to $B_{n}(x)$, the critical temperature is also affected to some degree. If we fix $C_{n}(x)=1$, the critical temperature is 159 (MeV). The result in Fig.\ \ref{fig:B} is the case in which $C_{n}(x)$ and $A_{n}(x)$ are simultaneously near $1$ within $10^{-3}$ order.\par
Consequently, $C_{n}(x)=A_{n}(x)$ in the general form of the exact quark propagator is a reasonable approximation. Moreover, if (IIa) is more correct from the viewpoint of gauge invariance, the SDE with the Landau gauge and $A_{n}^{\prime}(x)=1$ is the simplest, reasonable approximation in the framework of the ladder approximation.\par

\section{Summary and discussion}
The improved ladder approximation SDE at finite temperature has a gauge choice problem. For this problem, we calculated the SDE with the gauge parameter depending on an external momentum. By this method, the WTI is satisfied. Then, we used two cases for the exact quark propagator form, that is, $C_{n}(x)=A_{n}(x)$ and $C_{n}(x)\neq A_{n}(x)$.\par
The result with the Landau gauge for the critical temperature shows that $C_{n}(x)=A_{n}(x)$ is a reasonable approximation for the general form $C_{n}(x)\neq A_{n}(x)$. Thus, when one studies the critical temperature using the SDE, $C_{n}(x)=A_{n}(x)$ is the valid method.\par
To solve the SDE with the functional gauge parameter, we found that the functional gauge parameter has essentially large values for, e.g., $n=-2\sim 1$ and small $x (<\Lambda_{qcd})$. Thus, we expect that it is valid to choose the functional gauge parameter as a step function depending on the external momentum.\par
In the case with the functional gauge parameter, the critical temperature is nearly the Landau gauge with $A_{n}^{\prime}(x)=1$. For this reason, we found that the Landau gauge with $A_{n}^{\prime}(x)=1$ is a reasonable approximation from the viewpoint of the WTI.\par
Finally, we point out the uncertain part with respect to thermal effects in our calculational procedure. $C_{n}(x)$ and $A_{n}(x)$ include thermal effects. In particular, for real time, $C(p)\neq A(p)$ ($p_{0}$ is continuous) has a physical meaning. This generates characteristic collective modes, plasminos (for fermion). Thus, rigorous $C(p)=A(p)=1$ might have a problem. Therefore, for real time, it is difficult to satisfy the ladder approximation WTI fully. In view of this, $C_{n}(x)\simeq A_{n}(x)$ by numerical method in our calculation might be valid as an approximation.\par
On the other hand, we assume that this problem might not be critical in our calculational procedure with imaginary time, because the analytic continuation is necessary to change from imaginary time to real time, after performing the summation. Moreover, the relation between imaginary time and real time is not simple. For example, as shown in Ref.\ 7), there is an extra term in the SDE with real time. Hence, our calculational procedure with imaginary time should not provide correct results in real time. If we can make $C_{n}(x)=A_{n}(x)=1$ from our calculational procedure, it should become $C(p)=A(p)=1$ after the analytic continuation. However, we do not know whether this should be meaningful in real time. At least, we must study the existence of a solution corresponding to $C(p)=A(p)=1$ after the analytic continuation. In addition, since the functional gauge parameter that we used here depends on an external momentum, we must modify the functional gauge parameter in some way in real time. Therefore, we hypothesize that our method does not affect the existence of plasminos directly.\par
To understand the exact details, we should study the relation between real time and imaginary time by the same method used in Ref.\ 7).

%\section*{Acknowledgements}
%We would like to thank ...........

\appendix
\section{$L, I$ and $H$ in the SDE}
We show explicit expressions of $L,\ I$, and $H$ in Eqs.\ (\ref{eq:A1}), (\ref{eq:gene1}), and (\ref{eq:gene2}).

\[
a_{+}=(p_{0}-q_{0})^{2}-(x+y)^{2}\ ,\ a_{-}=(p_{0}-q_{0})^{2}-(x-y)^{2}.
\]
\begin{itemize}
\item$A_{n}^{\prime}(x)$

\[
L_{1}=-\frac{x^{2}+y^{2}-p_{0}^{2}-q_{0}^{2}}{2}\log\frac{a_{+}}{a_{-}}+2xy,
\]
\[
L_{2}=2xy-\frac{(p_{0}^{2}-q_{0}^{2}-x^{2}+y^{2})^{2}}{2}\Big(\frac{1}{a_{+}}-\frac{1}{a_{-}}\Big).
\]

\item$C_{n}(x)$

\[
I_{1}=2q_{0}\log\frac{a_{+}}{a_{-}}\ ,\ I_{2}=q_{0}\Big(-2(p_{0}-q_{0})^{2}\Big(\frac{1}{a_{+}}-\frac{1}{a_{-}}\Big)-\log\frac{a_{+}}{a_{-}}\Big),
\]

\[
I_{3}=-(p_{0}-q_{0})\Big[\log\frac{a_{+}}{a_{-}}-\Big(-(p_{0}-q_{0})^{2}+x^{2}-y^{2})\Big(\frac{1}{a_{+}}-\frac{1}{a_{-}}\Big)\Big].
\]

\item$A_{n}(x)$

\[
H_{1}=(p_{0}-q_{0})q_{0}\Big[(x^{2}-y^{2}+(p_{0}-q_{0})^{2})\Big(\frac{1}{a_{+}}-\frac{1}{a_{-}}\Big)+\log\frac{a_{+}}{a_{-}}\Big],
\]

\[
H_{2}=-4xy+(x^{2}+y^{2}-(p_{0}-q_{0})^{2})\log\frac{a_{+}}{a_{-}},
\]
\begin{align*}
H_{3}=\displaystyle \Big(x^{2}+y^{2}-&\displaystyle \frac{x^{2}+y^{2}-(p_{0}-q_{0})^{2}}{2}\Big)\log\frac{a_{+}}{a_{-}}\\[0.28cm]
&-\displaystyle \Big(\frac{(x^{2}-y^{2})^{2}-(p_{0}-q_{0})^{4}}{2}\Big)\Big(\frac{1}{a_{+}}-\frac{1}{a_{-}}\Big).
\end{align*}
\end{itemize}

\end{document}